\documentclass{PoS}

\newcommand{\beq}[1]{
\begin{equation}\label{#1}}
\newcommand{\eeq}{\end{equation}}
\newcommand{\bea}[1]{
\begin{eqnarray}\label{#1}}
\newcommand{\eea}{\end{eqnarray}}

\title{NLO BFKL at work: the electroproduction of two light vector mesons}

\ShortTitle{Electroproduction of two light vector mesons}

\author{\speaker{Alessandro Papa}\\
        Dipartimento di Fisica, Universit\`a della Calabria, 
        and INFN - Gruppo Collegato di Cosenza, \\I-87036 Rende, Italy\\
        E-mail: \email{papa@cs.infn.it}}

\author{Dmitry Yu. Ivanov\\
        Sobolev Institute of Mathematics, RU-630090 Novosibirsk, Russia\\
        E-mail: \email{d-ivanov@math.nsc.ru}}

\abstract{The amplitude for the forward electroproduction of two light vector mesons 
can be written completely within perturbative QCD in the Regge limit with next-to-leading 
accuracy, thus providing the first example of a physical application of the BFKL approach
at the next-to-leading order. Recently, a numerical determination of the amplitude has been
obtained in the case of equal photon virtualities, by using a definite representation for 
the amplitude and a definite optimization method for the perturbative series. Here, we 
study the main systematic effects in the previous determination, by considering a 
different representation of the amplitude and different optimization methods 
of the perturbative series. Moreover, we compare our result for the differential cross 
section at the minimum $|t|$ with a different approach, based on collinear
kernel improvement.}

\FullConference{DIFFRACTION 2006 - International Workshop on Diffraction in High-Energy Physics \\
		 September 5-10 2006\\
		 Adamantas, Milos island, Greece}

\begin{document}

\section{Introduction}

\begin{figure}
\centering
\includegraphics[width=0.5\textwidth]{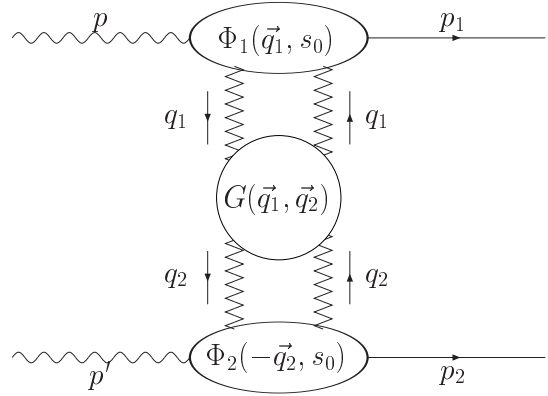}
\caption[]{Schematic representation of the amplitude for the $\gamma^*(p)\,
\gamma^*(p') \to V(p_1)\, V(p_2)$ scattering.}
\label{fig:BFKL}
\end{figure}

In the BFKL approach~\cite{BFKL}, both in the leading logarithmic approximation (LLA), 
which means resummation of all terms $(\alpha_s\ln(s))^n$, and in the next-to-leading 
approximation (NLA), which means resummation of all terms $\alpha_s(\alpha_s\ln(s))^n$, the 
(imaginary part of the) amplitude for a large-$s$ hard collision process can be 
written as the convolution of the Green's function of two interacting Reggeized 
gluons with the impact factors of the colliding particles (see, for example, Fig.~\ref{fig:BFKL}).

The Green's function is determined through the BFKL equation. The kernel of the BFKL equation
for singlet color representation, i.e. in the Pomeron channel, is known now both in the 
forward~\cite{NLA-kernel} and in the non-forward~\cite{FF05} cases.
On the other side, impact factors are known with NLA accuracy in a few cases:
colliding partons~\cite{partonIF}, forward jet production~\cite{BCV03}
and forward transition from a virtual photon $\gamma^*$ to a light neutral vector meson 
$V=\rho^0, \omega, \phi$~\cite{IKP04}. The most important impact factor 
for phenomenology, the $\gamma^* \to \gamma^*$ impact factor, is calling for 
a rather long calculation, which seems to be close to completion now~\cite{gammaIF,Cha}.

The $\gamma^* \to V$ forward impact factor can be used together with the NLA BFKL forward
Green's function to build, completely within perturbative QCD and with NLA accuracy, the 
amplitude of the $\gamma^* \gamma^* \to V V$ reaction. This amplitude provides us 
with an ideal theoretical laboratory for the investigation of several open
questions in the BFKL approach and for the comparison with different approaches.

In Ref.~\cite{IP06} it was shown how the $\gamma^* \to V$ impact factors and the BFKL 
Green's function can be put together to build up the NLA forward amplitude of the 
$\gamma^* \gamma^* \to V V$ process in the $\overline {\mbox{MS}}$ scheme and 
a convenient series representation for this amplitude was presented. Then, in the case of 
equal photon virtualities, i.e. in the so-called ``pure'' BFKL regime, a numerical 
study was carried out which led to conclude that the NLA corrections are large and 
of opposite sign with respect to the leading order and that they are dominated,
at the lower energies, by the NLA correction from impact factors. However,
a smooth behaviour of the (imaginary part of the) amplitude with the energy could be
nevertheless obtained, by optimizing the choice of the energy scale $s_0$ in the BFKL 
approach and of the renormalization scale $\mu_R$ which appear both in subleading terms. 
The optimization method adopted there was an 
adaptation of the ``principle of minimum sensitivity'' (PMS)~\cite{Stevenson} 
to the case where two energy parameters are present.

Here, we want to study the main systematic effects in the determination of Ref.~\cite{IP06}, 
by considering a different representation of the amplitude and by adopting 
different optimization methods of the perturbative series. 
Concerning the first effect, we consider here a representation of the NLA amplitude 
where almost all the NLA corrections coming from the kernel are
exponentiated. As for the second effect, we compare here the PMS optimization method 
with two other well-known methods of optimization of the perturbative series, namely
the fast apparent convergence (FAC) method~\cite{Grun} and the 
Brodsky-Lepage-Mackenzie (BLM) method~\cite{BLM}.

Finally, we compare some of our results with those of Ref.~\cite{EPSW1}, where
the same process has been considered using some version of a collinear kernel improvement.
A systematic study of the effect of collinear kernel improvement~\cite{Sal98,Ciafaloni,
Altarelli,Thorne,agustin,Peschanski} for the amplitude in question is in progress.

\section{Representations of the NLA amplitude}

The process under consideration is the production of two light vector 
mesons ($V=\rho^0, \omega, \phi$) in the collision of two virtual photons,
$ \gamma^*(p) \: \gamma^*(p')\to V(p_1) \:V(p_2)$.
Here, neglecting the meson mass $m_V$, $p_1$ and $p_2$ are taken as Sudakov vectors satisfying 
$p_1^2=p_2^2=0$ and $2(p_1 p_2)=s$; the virtual photon momenta are instead
$p =\alpha p_1-Q_1^2/(\alpha s) p_2$ and $p'=\alpha^\prime p_2-Q_2^2/(\alpha^\prime s) p_1$,
so that the photon virtualities turn to be $p^2=-Q_1^2$ and $(p')^2=-Q_2^2$.
We consider the kinematics when $ s\gg Q^2_{1,2}\gg \Lambda^2_{QCD}$ and
$\alpha=1+Q_2^2/s+{\cal O}(s^{-2})$, $\alpha^\prime =1+Q_1^2/s+{\cal O}(s^{-2})$.
In this case vector mesons are produced by longitudinally polarized photons in
the longitudinally polarized state~\cite{IKP04}. Other helicity amplitudes are
power suppressed, with a suppression factor $\sim m_V/Q_{1,2}$.
We will discuss here the amplitude of the forward scattering, i.e.
when the transverse momenta of produced $V$ mesons are zero or 
when the variable $t=(p_1-p)^2$ takes its maximal value $t_0=-Q_1^2Q_2^2/s+{\cal
O}(s^{-2})$.

In Ref.~\cite{IP06} the NLA forward amplitude has been written
as a spectral decomposition on the basis of eigenfunctions of the 
LLA BFKL kernel:
\[
\frac{\mbox{Im}_s\left( {\cal A} \right)}{D_1D_2}=\frac{s}{(2\pi)^2}
\int\limits^{+\infty}_{-\infty}
d\nu \left(\frac{s}{s_0}\right)^{\bar \alpha_s(\mu_R) \chi(\nu)}
\alpha_s^2(\mu_R) c_1(\nu)c_2(\nu)\left[1+\bar \alpha_s(\mu_R)
\left(\frac{c^{(1)}_1(\nu)}{c_1(\nu)}
+\frac{c^{(1)}_2(\nu)}{c_2(\nu)}\right)
\right.
\]
\beq{amplnla}
\left.
+\bar \alpha_s^2(\mu_R)\ln\left(\frac{s}{s_0}\right)
\left(\bar
\chi(\nu)+\frac{\beta_0}{8N_c}\chi(\nu)\left[-\chi(\nu)+\frac{10}{3}
+i\frac{d\ln(\frac{c_1(\nu)}{c_2(\nu)})}{d\nu}+2\ln(\mu_R^2)\right]
\right)\right] \; .
\eeq
Here, ${\bar \alpha_s}=\alpha_s N_c/\pi$ and $D_{1,2}=-4\pi e_q  f_V/(N_c Q_{1,2})$,
where $f_V$ is the meson dimensional coupling constant ($f_{\rho}\approx
200\, \rm{ MeV}$) and $e_q$ should be replaced by $e/\sqrt{2}$, $e/(3\sqrt{2})$
and $-e/3$ for the case of $\rho^0$, $\omega$ and $\phi$ meson production,
respectively. We refer to Ref.~\cite{IP06} for the details of the derivation and for the 
definition of the functions of $\nu$ entering this expression.
Two energy scales enter the expression~(\ref{amplnla}), the renormalization scale $\mu_R$ 
and the scale $s_0$, which is an artificial scale introduced in the BFKL approach at the time
to perform the Mellin transform from the $s$-space to the complex angular
momentum plane.

It is easy to see that the above expression can be organized as a series:
\bea{series}
\frac{Q_1Q_2}{D_1 D_2}\frac{\mbox{Im}_s ({\cal A}_{\mathrm series})}{s} &=&
\frac{1}{(2\pi)^2}  \alpha_s(\mu_R)^2 \label{honest_NLA} \\
& \times &
\biggl[ b_0
+\sum_{n=1}^{\infty}\bar \alpha_s(\mu_R)^n   \, b_n \,
\biggl(\ln\left(\frac{s}{s_0}\right)^n   +
d_n(s_0,\mu_R)\ln\left(\frac{s}{s_0}\right)^{n-1}     \biggr)
\biggr]\;. \nonumber
\eea
The $b_n$ coefficients are determined by the kernel and the impact factors 
in LLA, while the $d_n$ coefficients depend also on the NLA corrections to the kernel and 
to the impact factors. For their expression, see Ref.~\cite{IP06}. 

An alternative possibility to represent the NLA amplitude is obtained by 
exponentiating the bulk of the kernel NLA corrections, 
\[
\frac{\mbox{Im}_s\left( {\cal A}_{\mathrm exp}\right)}{D_1D_2}
=\frac{s}{(2\pi)^2}
\int\limits^{+\infty}_{-\infty}
d\nu \left(\frac{s}{s_0}\right)^{\bar \alpha_s(\mu_R)
\chi(\nu)+\bar \alpha_s^2(\mu_R)
\left(
\bar
\chi(\nu)+\frac{\beta_0}{8N_c}\chi(\nu)\left[-\chi(\nu)+\frac{10}{3}
\right]
\right)}
\alpha_s^2(\mu_R) c_1(\nu)c_2(\nu)
\]
\beq{amplnlaE}
\times\! \left[1+\bar \alpha_s(\mu_R)
\left(\frac{c^{(1)}_1(\nu)}{c_1(\nu)}
+\frac{c^{(1)}_2(\nu)}{c_2(\nu)}\right)
+\bar \alpha_s^2(\mu_R)\ln\left(\frac{s}{s_0}\right)
\frac{\beta_0}{8N_c}\chi(\nu)\left(
i\frac{d\ln(\frac{c_1(\nu)}{c_2(\nu)})}{d\nu}+2\ln(\mu_R^2)
\right)\right].
\eeq
This form of the NLA amplitude was used in~\cite{KIM1} (see also~\cite{KIM2}), 
without account of the last two terms in the second line of (\ref{amplnlaE}), 
for the analysis of the total $\gamma^*\gamma^*$ cross section. We will
refer in the following to this representation simply as ``exponentiated''
amplitude.

It is easily seen that the amplitude, in any of the given representations, is independent 
in the NLA from the choice of $s_0$ and of $\mu_R$~\cite{IP06}.

\section{Numerical results}

\begin{figure}
\includegraphics[width=0.49\textwidth]{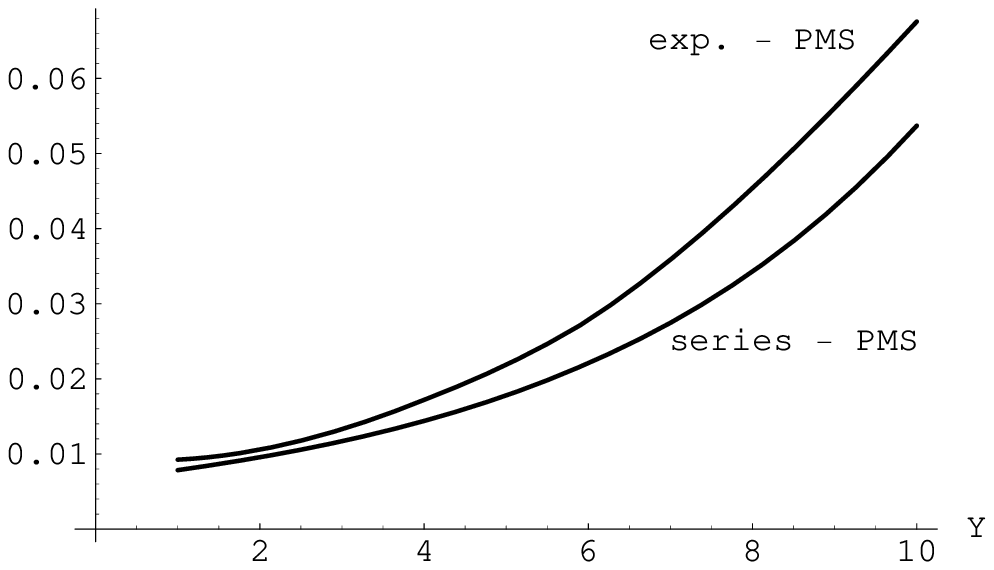}
\includegraphics[width=0.49\textwidth]{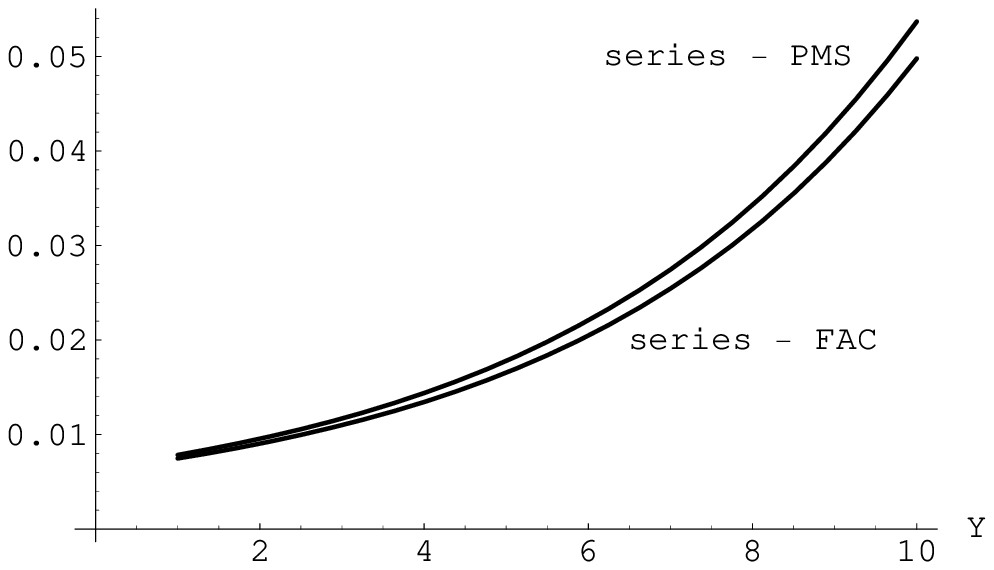}
\caption[]{$\mbox{Im}_s ({\cal A})Q^2/(s \, D_1 D_2)$ as a function of $Y$ 
at $Q^2$=24 GeV$^2$ ($n_f=5$): (left) series representation with PMS and
``exponentiated'' representation with PMS, (right) series representation with PMS and
with FAC.}
\label{comp1}
\end{figure}

\begin{figure}
\includegraphics[width=0.49\textwidth]{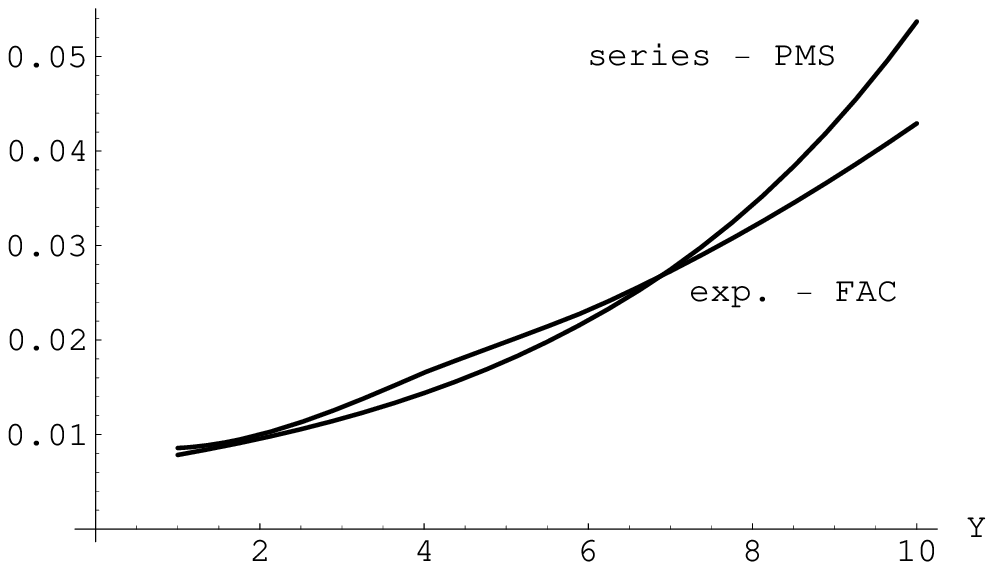}
\includegraphics[width=0.49\textwidth]{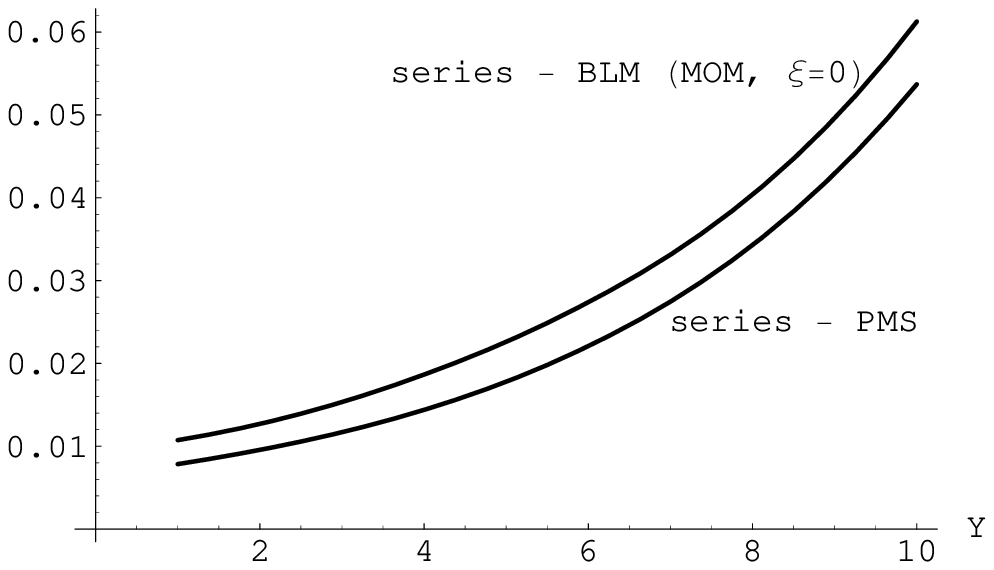}
\caption[]{$\mbox{Im}_s ({\cal A})Q^2/(s \, D_1 D_2)$ as a function of
$Y$ at $Q^2$=24 GeV$^2$ ($n_f=5$): (left) series representation with PMS 
and ``exponentiated'' representation with FAC, (right) series representation with PMS and 
with BLM.}
\label{comp2}
\end{figure}

In Ref.~\cite{IP06} we presented some numerical results for the 
amplitude given in Eq.~(\ref{series}) for the $Q_1=Q_2\equiv Q$ kinematics, 
i.e. in the ``pure'' BFKL regime. We found that the $d_n$ coefficients are negative 
and increasingly large in absolute values as the perturbative order increases, making
evident the need of an optimization of the perturbative series. We adopted the principle 
of minimal sensitivity (PMS)~\cite{Stevenson}, by requiring the minimal sensitivity of 
the predictions to the change of both the renormalization and the energy scales, 
$\mu_R$ and $s_0$. We considered the amplitude for 
$Q^2$=24 GeV$^2$ and $n_f=5$ and studied its sensitivity to variation
of the parameters $\mu_R$ and $Y_0=\ln(s_0/Q^2)$. We could see that for
each value of $Y=\ln(s/Q^2)$ there are quite large regions in
$\mu_R$ and $Y_0$ where the amplitude is practically independent on $\mu_R$ and
$Y_0$ and we got for the amplitude a smooth behaviour in $Y$ (see the curve
labeled ``series - PMS'' in Figs.~\ref{comp1} and~\ref{comp2}).
The optimal values turned out to be $\mu_R\simeq 10 Q$
and $Y_0\simeq 2$, quite far from the kinematical values $\mu_R=Q$
and $Y_0=0$. These ``unnatural'' values probably mimic large unknown
NNLA corrections.


As an estimation of the systematic effects in our determination, we want
to consider here also the ``exponentiated'' representation of the amplitude,
Eq.~(\ref{amplnlaE}), and different optimization methods.\footnote{For more details
on the following, see Ref.~\cite{IP}.}

At first, we compare the series and the ``exponentiated'' determinations
using in both case the PMS method. The optimal values of $\mu_R$ and $Y_0$
for the ``exponentiated'' amplitude are quite similar to those obtained in the case 
of the series representation, with only a slight decrease of the optimal $\mu_R$. 
Fig.~\ref{comp1} (left) shows that the two determinations are 
in good agreement at the lower energies, but deviate increasingly for large
values of $Y$. It should be stressed, however, that the applicability
domain of the BFKL approach is determined by the condition 
$\bar \alpha_s(\mu_R) Y \sim 1$ and, for $Q^2$=24 GeV$^2$ and for the 
typical optimal values of $\mu_R$, one gets from this condition $Y\sim 5$. 
Around this value the discrepancy between the two determinations is within a 
few percent.

As a second check, we changed the optimization method and applied
it both to the series and to the ``exponentiated'' representation.
The method considered is the fast apparent convergence (FAC) 
method~\cite{Grun}, whose strategy, when applied to a usual perturbative
expansion, is to fix the renormalization scale to the value for which the
highest order correction term is exactly zero. In our case, the application
of the FAC method requires an adaptation, for two reasons: the first is that
we have two energy parameters in the game, $\mu_R$ and $Y_0$, the second
is that, if only strict NLA corrections are taken, the amplitude 
does not depend at all on these parameters. For details about the
application of this method, we refer to~\cite{IP}. Here, we merely show
the results: the FAC method applied to the series representation (see Fig.~\ref{comp1} (right))
and to the exponentiated representation (see Fig.~\ref{comp2} (left)) gives results in
nice agreement with those from the PMS method applied to the series representation, over 
the whole energy range considered. 


Another popular optimization method is the Brodsky-Lepage-Mackenzie (BLM)
one~\cite{BLM}, which amounts to perform a finite renormalization
to a physical scheme and then to choose the renormalization scale in order to
remove the $\beta_0$-dependent part. We applied this method only to the 
series representation, Eq.~(\ref{series}). The result is compared with the
PMS method in Fig.~\ref{comp2} (right) (for details, see Ref.~\cite{IP}).

%

The $\gamma^* \gamma^* \to \rho \rho $ amplitude with the inclusion
of NLA BFKL effects has been studied also in Ref.~\cite{EPSW1}. In that paper,
the amplitude has been built with the following ingredients: leading-order 
impact factors for the $\gamma^* \to \rho$ transition, BLM scale fixing 
for the running of the coupling in the prefactor of the amplitude (the BLM 
scale is found using the NLA $\gamma^* \to \rho$ impact factor 
calculated in Ref.~\cite{IKP04}) and renormalization-group-resummed 
BFKL kernel, with resummation performed on the LLA BFKL kernel at fixed 
coupling~\cite{KMRS04}. In Ref.~\cite{EPSW1} the behaviour of $d\sigma/dt$ at $t=t_0$ 
as a function of $\sqrt{s}$ was determined for three values of
the common photon virtuality, $Q$=2, 3 and 4 GeV. 

\begin{figure}
\includegraphics[width=0.49\textwidth]{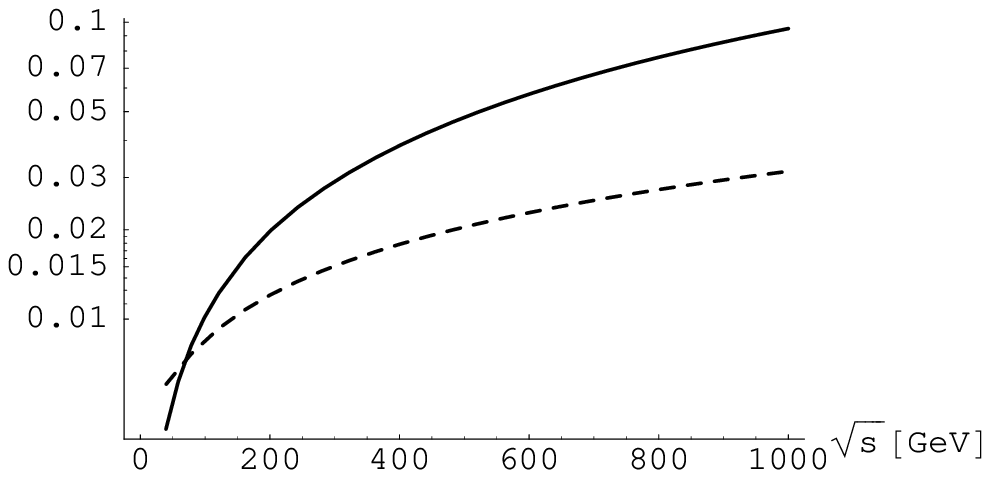}
\includegraphics[width=0.49\textwidth]{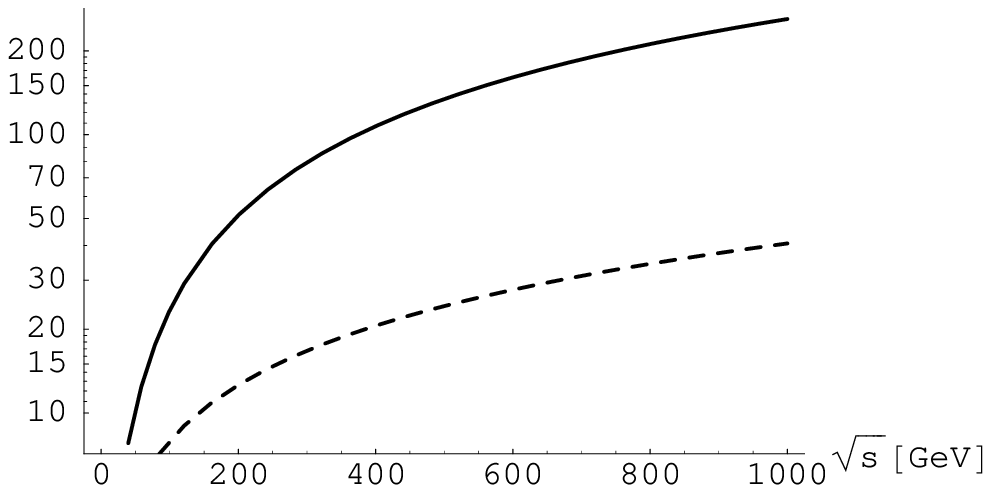}
\caption[]{$\left. d\sigma/dt\right|_{t=t_0}$ [pb/GeV$^2$] as a function of
$\sqrt{s}$ at $Q^2$=16 GeV$^2$ ($n_f=5$) (left) and at $Q^2$=4 GeV$^2$ ($n_f=3$) (right)
from the series representation with the PMS optimization method (solid lines) compared 
with the determination from the approach in Ref.~\cite{EPSW1} (dashed lines).}
\label{dsig}
\end{figure}

In order to make a comparison with the findings of Ref.~\cite{EPSW1},
we computed $d\sigma/dt$ at $t=t_0$ for $Q$=2 and $Q$=4 GeV as functions
of $\sqrt{s}$. We used $f_\rho$=216 MeV, $\alpha_\mathrm{EM}=1/137$ and
the two--loop running strong coupling corresponding to the value 
$\alpha_s(M_Z)=0.12$. The results are shown in the linear-log
plots of Fig.~\ref{dsig}, which shows a large 
disagreement. It would be interesting to understand to what extent 
this disagreement is due to the use in Ref.~\cite{EPSW1} of  
LLA impact factors instead of the NLA ones or to the way the collinear
improvement of the kernel is performed.

\vspace{0.5cm}

The work of D.I. was partially supported by grants RFBR-05-02-1611, NSh-5362.2006.2.

\end{document}